\documentclass[apj]{emulateapj}
\usepackage{apjfonts}

\slugcomment{Accepted for publication in The Astrophysical Journal}
\shorttitle{CONVECTION DURING SIMMERING} 
\shortauthors{PIRO \& CHANG}

\newcommand{\be}{\begin{eqnarray}}
\newcommand{\ee}{\end{eqnarray}}
\newcommand{\lp}{\left(}
\newcommand{\rp}{\right)}

\begin{document}


\title{Convection during the Late Stages of Simmering in Type Ia Supernovae}

\author{Anthony L. Piro and Philip Chang\altaffilmark{1}}

\affil{Astronomy Department and Theoretical Astrophysics Center,
University of California,
Berkeley, CA 94720;\\ tpiro@astro.berkeley.edu, pchang@astro.berkeley.edu}

\altaffiltext{1}{Miller Institute for Basic Research, University of California, Berkeley, CA 
94720.}


\begin{abstract}
Following unstable ignition of carbon, but prior to explosion, a white dwarf (WD) in a Type Ia supernova (SN Ia) undergoes a simmering phase. During this time, a central convective region grows and encompasses $\sim1M_\odot$ of the WD over a timescale of $\sim 10^3$ yrs, which sets the thermal and turbulent profile for the subsequent explosion. We study this time-dependent convection and summarize some of the key features that differ from the traditional, steady-state case. We show that the long conductive timescale above the convective zone and the extraction of energy to heat the WD core leads to a decrease of the convective luminosity and characteristic velocities near the convective zone's top boundary.  In addition, differences in the composition between the convective core and the conductive exterior will significantly alter the location of this boundary. In this respect, we find the biggest effect due to complete $^{22}$Ne sedimentation prior to carbon ignition. These effects add diversity to the possible WD models, which may alter the properties of the SN Ia explosion.

\end{abstract}



\section{Introduction}

   The use of Type Ia supernovae (SNe Ia) as cosmic distance indicators
has focused attention to the study of white dwarf (WD) explosions. Of particular
importance is determining the parameters that dictate the observed
diversity of SNe Ia. Recent modeling
demonstrates that the variation along the width-luminosity relation
\citep{phi99} may be explained by large  variations in the abundance of stable
iron group elements \citep{kw07,woo07}
with the dominant cause for  diversity likely residing in the explosion mechanism
\citep{maz07}. Another important variable is the metallicity of the WD core \citep{tbt03}.

   It is critical to explore the initial conditions that may add to the diversity.
Simulations of turbulent thermonuclear flames in WDs
have demonstrated that  the composition and energy of ejecta depend sensitively
on the competition between flame propagation, instabilities driven by
turbulence, and expansion of the WD \citep[][and references therein]{hn00}. 
These simulations should therefore depend on the thermal and
turbulent state of the WD set by the pre-explosive convective simmering
phase \citep{nom84,ww86}.

   During the early stages of simmering many studies have focused on
the convective Urca process
\citep{pac72,bru73,ca75,ibe78a,ibe78b,ibe82,bw90,moc96,ste99,bk01,les05,sw06}.
This occurs when nuclei repeatedly electron capture and beta decay as they are carried
by convection back and forth across the electron capture threshold density (the ``Urca shell'').
Although the energy loss from this process is most likely not great enough to cause
global cooling, it can still have a significant effect on the convective motions \citep{les05}.
Once the central temperature has grown above $\approx(5-6)\times10^8\ {\rm K}$
(which corresponds to $\sim10^5\ {\rm s}$ before the burning wave begins),
there is no longer time for electron captures on $^{23}$Na \citep{pb07}, and the
convective Urca process will cease. During the last $\sim10^5\ {\rm s}$
any compositional gradients are mixed homogeneously
by subsequent convection.

   An additional place where simmering is important is for understanding the
conditions within the WD immediately prior to the explosion \citep{gw95,hs02,woo04,ww04}.
The properties of the temperature fluctuations present in the
convection set the size and distribution of the ignition points, which are
crucial for determining the success of the subsequent burning wave
\citep[see][and references therein]{rop06}. The interaction of convection with
rotation sets the morphology of convective motions \citep{kwg06} as well as the
overall rotation profile of the WD \citep{pir08}.

   The last way simmering has gained attention is in its
ability to enhance the neutron abundance in the WD core \citep{pb07,cha07}. This happens primarily
via the reaction chain $^{12}$C$(p,\gamma)^{13}$N$(e^-,\nu_e)^{13}$C,
where the protons are leftover from $^{12}$C burning. Depending on the amount of
carbon that is consumed before burning becomes dynamical, as well as the density at which
it takes place, this neutronization enhancement could very well be large enough to
mask any trend expected with metallicity in environments that have roughly sub-solar
metallicity.

   In this present work we focus on the general properties of the simmering convection, with
the aim of identifying characteristics that may introduce diversity to the SN Ia progenitors.
We begin in \S \ref{sec:models} by presenting the main features of
our models. We illustrate how time-dependent convection in the simmering phase differs
from the familiar case of steady-state convection. In this new picture, the convective flux decreases
outside the central heating zone due to both the heating of new material as the convective region
grows and the inability to transfer significant energy to the conductive exterior.
In \S \ref{sec:boundary} we explore the location of the top of the convective zone.
We point out that degeneracy effects enhance the response of the boundary location
to changes in composition. We conclude with a summary of our results and a discussion
of future work in \S \ref{sec:conclusion}.


\section{Luminosity and Characteristic Velocities\\ for Expanding Convection}
\label{sec:models}

   We begin by summarizing the main features of our simmering models.
(For further details, the interested reader
should refer to Woosley et al. 2004; Lesaffre et al.~2006; Piro \& Bildsten 2007; Piro 2008.)
Unstable ignition of $^{12}$C occurs when the heating from carbon fusion
beats neutrino cooling. The central temperature then rises and
a convective zone grows outward, eventually
encompassing $\sim1 M_\odot$ of the WD after $\sim10^3\ {\rm yrs}$.
As the central temperature, $T_c$, increases, carbon burning becomes more
vigorous and the heating timescale, $t_h\equiv (d\ln T_c/dt)^{-1}$, gets shorter. This timescale
in general depends on the size of the region responding to the rising central
temperature at the core.

   Simmering ends and a burning wave commences once $t_h\lesssim t_{\rm conv}$, where
$t_{\rm conv}$ is eddy overturn timescale. At these late times, individual eddies
may experience significant heating during their transit \citep{gw95}, so that the temperature
profile is no longer an adiabat and the entire convective core does not respond to the increasing
central temperature. (In contrast, we show below that during the majority of the time
$t_h$ depends on the heat capacity of the entire convective mass.)
This makes it difficult to exactly calculate the {\it precise} moment
when simmering ends. For this reason
\citet{les06} explore $t_{\rm conv}=\alpha t_h$, where $\alpha~\lesssim~1$
parameterizes this uncertainty. Since we are only roughly concerned with
resolving the end of the simmering phase, we take
$t_{\rm conv}\approx t_h\approx c_p T_c/\epsilon$, where $c_p$ is the specific heat
capacity at constant pressure, $\epsilon$ is the heating from carbon burning,
and all these quantities are evaluated at the WD center.
This estimates that simmering should end when $t_h\approx7\ {\rm s}$ at a central temperature and
density of $T_c\approx7.8\times10^8\ {\rm K}$
and $\rho_c\approx2.6\times10^9\ {\rm g\ cm^{-3}}$, which is roughly in agreement with
the results presented by \citet{woo04} using the Kepler stellar evolution
code \citep{wea78}.

   We follow the simmering phase by calculating a series of hydrostatic WD models,
each with a different central temperature, but at a fixed mass \citep[see][]{pb07,pir08}.
For simplicity we ignore the convective Urca process since our focus is on
the last $\sim10^5\ {\rm s}$.
The timescale for thermal conduction across the WD
is $t_{\rm th}\equiv K_c/R^2\sim10^6\ {\rm yrs}$, where $K_c$ is the conductivity
and $R$ is the radius, which is much longer the timescale over which heating is occurring.
Therefore the
convection efficiently mixes entropy and the convective region nearly follows an adiabat
out from the WD center.
Outside the convective zone, we assume the WD is
isothermal with a temperature $T_i$. 

\subsection{Convective Luminosity}
\label{sec:luminosity}

   To understand how time-dependent convection is different than from
that normally found in steady-state convection,
we focus on the time-dependent entropy equation
\be
	c_p\frac{\partial T}{\partial t} = \epsilon - \frac{\partial L_c}{\partial M_r},
	\label{eq:entropy}
\ee
where $L_c$ is the convective luminosity. This equation omits the work required to
expand the WD as the heating takes place, which is a significant amount of
energy and thus requires some discussion. For each convective model we compared the total
change in WD binding energy to the total change in internal energy of the electrons (which are primarily
degenerate and relativistic), including the ion-electron Coulomb interaction energy
\citep[according to][]{cp98}. These two quantities are equal to
the numerical accuracy of our integrations, which demonstrates that all of the work
required to expand the WD comes from changes in the internal energy of the electrons.
Thus, the entropy created from nuclear burning all
goes into convective motions or the internal thermal energy, and we are justified in omitting
the binding energy and electron internal energy terms from equation (\ref{eq:entropy}).

   The temperature profile in the convective zone follows an adiabat with a power law index
$n\equiv(\partial\ln T/\partial\ln P)_{\rm ad}$. The time derivative
of the temperature at a given pressure can be expressed as
\be
	\partial T/\partial t& =& (P/P_c)^n\left[ \partial T_c/\partial t +T_c\ln(P/P_c)\partial n/\partial t\right]
	\nonumber
	\\
	&\approx& (P/P_c)^n \partial T_c/\partial t,
\ee
where $P_c$ is the central pressure,
and for simplicity we are assuming that it does not change
appreciably in time. From this we see that the timescale for the temperature
change {\it at any pressure} is set by the central temperature change
\be
	\frac{\partial\ln T}{\partial t} = \frac{\partial \ln T_c}{\partial t}\equiv \frac{1}{t_h}.
	\label{eq:tgr}
\ee
Therefore there is a well-defined, global heating timescale,
$t_h$\footnote{Note that this timescale is different than the local timescale,
$c_pT_c/\epsilon$, used above for estimating the end of
simmering. This is because, with the exception of late times, the convective zone
is well-coupled.}.
To account for the changing central pressure in a more rigorous calculation,
we must take partial derivatives at constant mass coordinate, $M_r$. This can be
performed by inverting the empirically found $M_c(T_i,T_c)$ relation presented in
Piro (2008; or see eq. [\ref{eq:mct}] below), where $T_i$ is the nearly isothermal
temperature of the non-convective, conductive region. The result is
\be
	T(T_c,M_r) = 0.83T_c\left[ 1-\lp\frac{\mu_e}{2}\rp^2\frac{M_r}{1.48\ M_\odot}\right],
	\label{eq:tmr}
\ee
where $\mu_e$ is the mean molecular weight per electron.
This can be used to find $(d\ln T/dt)_{M_r}=d\ln T_c/dt$, which confirms
our conclusion that $t_h$ is the same at any depth within the convective zone.

   Multiplying equation (\ref{eq:entropy}) by $dM_r=4\pi r^2\rho dr$ and integrating,
\be
	\int_0^{M_r}c_p\frac{\partial T}{\partial t}dM_r = \int_0^{M_r}\epsilon dM_r
		- L_c(M_r)+L_c(0).
	\label{eq:integrated}
\ee
We pull $t_h$ outside of the left-hand integral to find
\be
	\int_0^{M_r}c_pT\frac{\partial\ln T}{\partial t}dM_r
	= \frac{1}{t_h}\int_0^{M_r}c_pTdM_r=\frac{E_{\rm th}(M_r)}{t_h},
\ee
where $E_{\rm th}(M_r)$ is the integrated thermal energy up to a mass coordinate $M_r$.
We set $L_c(0)=0$ and define the nuclear luminosity as $L_{\rm nuc}=\int\epsilon dM_r$,
so that equation (\ref{eq:integrated}) becomes
\be
	\frac{E_{\rm th}(M_r)}{t_h} = L_{\rm nuc}(M_r)-L_c(M_r)
	\label{eq:at_mr}
\ee
Equation (\ref{eq:at_mr}) expresses that the nuclear luminosity must either go into thermal heating
or convective motions, and it is valid at any $M_r$. It contains two unknowns,
$t_h$ and $L_c(M_r)$. We set the luminosity at the surface of the convective zone to be zero,
$L_c(M_c)=0$. This boundary is required since $t_{\rm th}$ is long in the non-convective
regions, which prevents significant heat transfer.
We can then solve for $t_h$,
\be
	t_h = E_{\rm th}(M_c)/L_{\rm nuc}(M_c),
	\label{eq:th}
\ee
which matches the definition of $t_h$ that \citet{wei06} use in the context
of type I X-ray bursts on neutron stars.
We substitute $t_h$ back into equation (\ref{eq:at_mr}) to get the convective luminosity
\be
	L_c(M_r) &=& L_{\rm nuc}(M_r)-E_{\rm th}(M_r)/t_h
	\nonumber
	\\
	&=& L_{\rm nuc}(M_r)\left[1-\frac{E_{\rm th}(M_r)}{E_{\rm th}(M_c)}
		\frac{L_{\rm nuc}(M_c)}{L_{\rm nuc}(M_r)}\right].
	\label{eq:correct}
\ee
For steady-state convection, $L_c(M_r)=L_{\rm nuc}(M_r)$. The ratio
$E_{\rm th}(M_r)/E_{\rm th}(M_c)$ is the modification due to the growing
nature of the convection and the $L_{\rm nuc}(M_c)/L_{\rm nuc}(M_r)$ term is from
long thermal time for the conductive exterior, which forces $L_c(M_c)=0$.

\begin{figure}
\epsscale{1.2}
\plotone{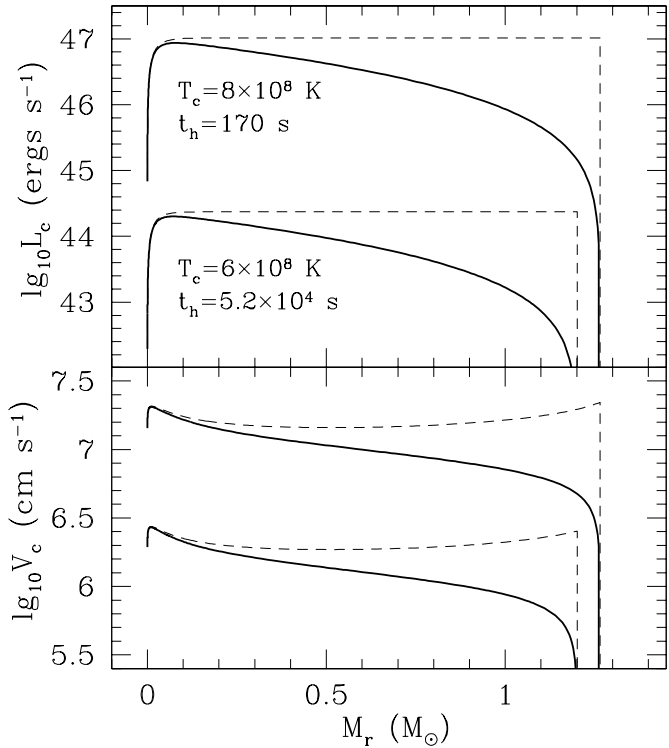}
\caption{The convective luminosity, $L_c$ and characteristic velocities, $V_c$, as
a function of mass coordinate, $M_r$. The upper (lower) lines in each panel are for
a central temperature of $T_c=8\times10^8\ {\rm K}$ ($6\times10^8\ {\rm K}$).
Solid lines are the estimates for time-dependent convection using eq. (\ref{eq:correct}).
The dashed lines are calculations that assume $L_c(M_r)=L_{\rm nuc}(M_r)$,
which are plotted for comparison.}
\label{fig:luminosity}
\epsscale{1.0}
\end{figure}

   In the upper panel of Figure \ref{fig:luminosity} we plot as solid lines the convective
luminosity found using equation (\ref{eq:correct}).  These models all have a composition
of 0.5 $^{12}$C, 0.48 $^{16}$O, and 0.02 $^{22}$Ne by mass fraction, with a mass of
$1.37\ M_\odot$ and an initial isothermal temperature $T_i=10^8\ {\rm K}$.
We solve for $\rho$ using the \citet{pac83} fit for the equation of state, and include the
Coulomb energy of \citet{cp98}.  We present central temperatures of
$T_c=6\times10^8$ and $8\times10^8\ {\rm K}$,
which corresponds to $t_h=14\ {\rm hrs}$ and $170\ {\rm s}$, respectively. The $t_h$ associated
with the latter case is an overestimate since, as mentioned above, at these late times during
simmering only inner portion of the core responds to the rising central temperature
(in effect decreasing $E_{\rm th}$ in eq. [\ref{eq:th}]).
The energy generation rate for $^{12}$C burning
is taken from \citet{cf88} with strong screening included from \citet{svh69}.
Also plotted in Figure \ref{fig:luminosity} is the convective luminosity for
$L_c(M_r)=L_{\rm nuc}(M_r)$, i.e., for steady-state convection (dashed lines).
Near the center, $E_{\rm th}(M_r)$ is small
and grows less quickly than $L_{\rm nuc}(M_r)$,
so $L_c(M_r)$ is initially $\approx L_{\rm nuc}(M_r)$ 
(see eq. [\ref{eq:correct}]). At larger $M_r$, $L_c$
decreases due to the effects we have highlighted.

\subsection{Convective Velocities}
\label{sec:velocities}

If we take a characteristic eddy scale $l_c$, the thermal conduction timescale across an eddy
is $\sim(l_c/R)^210^6\ {\rm yrs}\sim(H/R)^210^6\ {\rm yrs}\sim10^4\ {\rm yrs}$, where
$H$ is the pressure scale height. Since this
timescale is long, the convection is efficient \citep{hk94}.
Using estimates from mixing-length theory,
the characteristic convective velocity,
$V_{\rm conv}$, is related to $F_{\rm conv}$ via
\be
	V_c = \lp \frac{\mathcal{Q}gl_c}{c_pT}\frac{F_c}{\rho}\rp^{1/3}
	\sim \lp\frac{F_c}{\rho}\rp^{1/3},
\ee
where $\mathcal{Q}=-(\partial\ln\rho/\partial\ln T)_P$ and $g=GM_r/r^2$ is the
local gravitational acceleration, and we have set $l_c\approx H$.

   In the bottom panel of Figure \ref{fig:luminosity} we have plotted $V_c$,
setting the mixing-length to the scale height, 
$l_c=H$ ({\it solid lines}). The shape of these velocity profiles
are similar to those \citet{les05} present in the context of studying
the convective Urca process, which is active at much earlier times during
the simmering. The convective
velocities are different by as much as $\sim50\%$ from the na\"{i}ve estimate of
$L_c=L_{\rm nuc}$ ({\it dashed lines}) near the top of the convective zone.


\section{The Convective Boundary Location}
\label{sec:boundary}

   Buoyantly rising eddies ascend
until their density matches their surroundings. The boundary between the
convective and isothermal zones is therefore set by a neutral buoyancy condition.
In practice this means that both the pressure and density
must be continuous. If both the convective and isothermal
regions have the same composition, the boundary is simply set by when the
adiabatic temperature of the
convective zone reaches the isothermal temperature, $T_i$
\citep[i.e., where the entropy matches,][]{hs02}.
If the composition is different, a buoyantly rising eddy will be prevented from
passing very far into the isothermal region and truncates the size of the convective
zone. This creates an abrupt change in temperature at the boundary of size
$\Delta T$. In actuality this change will be smoothed by overshoot and mixing,
so that the entropy remains continuous \citep[see, for example,][]{kwg06}.
The thermal and compositional structure at this boundary will be complicated,
especially at the last moments of the simmering, when relatively few convective
overturns take place for any given location of the convective boundary. Since the
convection is always sub-sonic, we expect the overshoot to be modest and
the simplification of a sharp compositional boundary to be adequate to
estimate the size of the convective zone.

    \citet{pir08} presented an empirically
derived relationship for the convective
boundary, $M_c$, as a function of the ratio of the isothermal and central
temperatures, $T_i/T_c$,
\be
	M_c = 1.48\ M_\odot\lp\frac{2}{\mu_e}\rp^2\left[ 1-1.2\frac{T_i}{T_c}\right].
	\label{eq:mct}
\ee
We make the substitutions $M_c\rightarrow M_c+\Delta M_c$ and $T_i\rightarrow T_i+\Delta T$
to solve for the change in the convective mass due to a temperature discontinuity,
\be
	\Delta M_c = -0.18\ M_\odot\lp\frac{2}{\mu_e}\rp^2\lp\frac{8}{T_c/T_i}\rp\frac{\Delta T}{T_i},
	\label{eq:deltamc}
\ee
where we have scaled to a temperature ratio $T_c/T_i=8$, as is appropriate
for near the end of the simmering phase.

   The temperature discontinuity allows a small conductive wave to propagate out from the
top boundary of the convective zone. Since the thermal conduction timescale
($t_{\rm th}\sim10^6\ {\rm yrs}$) is long
in comparison to the heating timescale
($t_h\sim10\ {\rm s}-10^3\ {\rm yrs}$), this wave can only travel a distance
$\sim (t_h/t_{\rm th})^{1/2}H\ll H$ before the growing convective zone overtakes it.
For this reason, we ignore this detail in the following calculations.

   In the next sections we study the change in the convective boundary analytically.
We discuss two main ways in which compositional discontinuities can be important.
These are changes in the neutron abundance and changes in the composition, which affects
the Coulomb corrections to the equation of state. In \S \ref{sec:numerical} we compare
the models summarized in Table \ref{tab:models} numerically to confirm these analytic results.

\begin{deluxetable}{l l l c c c c}
  \tablecolumns{7} \tablewidth{0pt}
  \tablecaption{Compositional Summary for Numerical Models}

  \tablehead{
    \colhead{Model} & \colhead{Purpose} & \colhead{Zone} & \colhead{$X_{12}$} & \colhead{$X_{13}$} & \colhead{$X_{16}$}
    & \colhead{$X_{22}$} }
  \startdata
  1 & Homogeneous &Convection & 0.5 & 0.0 & 0.48 & 0.02\\
     & &Isothermal & 0.5 & 0.0 & 0.48 & 0.02\\ 
  2 & Neutronization &Convection & 0.48 & 0.02 & 0.48 & 0.02 \\
     & &Isothermal & 0.5 & 0.0 & 0.48 & 0.02 \\
  3 & Comp. Gradient &Convection & 0.5 & 0.0 & 0.5 & 0.0 \\
     & &Isothermal & 0.7 & 0.0 & 0.3 & 0.0\\
  4 & Sedimentation&Convection & 0.487 & 0.0 & 0.487 & 0.026\\
  & & Isothermal & 0.5 & 0.0 & 0.5 & 0.0
   \enddata
   \label{tab:models}
\end{deluxetable}


\subsection{Neutron Abundance Discontinuity}
\label{sec:neutronization}

   We first study differences in neutron abundance between the convective and
isothermal zones. The neutronization is typically expressed as
\be
	Y_e= \frac{1}{\mu_e} = \sum_i \frac{Z_i}{A_i}X_i,
\ee
where $A_i$ and $Z_i$ are the nucleon number and charge of species $i$
with mass fraction $X_i$. The initial metallicity of the SN Ia progenitor is determined by the isotope
$^{22}$Ne, which has two additional neutrons \citep{tbt03}.
A mass fraction $X_{22}$ of $^{22}$Ne decreases $Y_e$ by an amount
$\Delta Y_e=2X_{22}/22\approx1.8\times10^{-3}X_{22}/0.02$. A large enhancement
of $^{22}$Ne could be present in the convective core if substantial gravitational separation
has occurred \citep{bh01,db02,gar07}. Neutron enhancement in the convective zone can
also occur from electron captures during the simmering, which decreases
$Y_e$ by an amount $\Delta Y_e\sim10^{-4}-10^{-3}$
\citep{pb07,cha07}.

   We first consider the simplest equation of state where relativistic degenerate electrons
dominate the pressure with corrections from the ideal gas of ions
(in \S \ref{sec:nuclei} we consider Coulomb corrections),
\be
	P = K\lp\frac{\rho}{\mu_e}\rp^{4/3} + \frac{\rho k_{\rm B}T}{\mu_i m_p},
	\label{eq:eos}
\ee
where $K=1.231\times10^{15}\ {\rm cgs}$ and $\mu_i$ is the mean molecular weight
per ion.

   We set the temperature to $T_i+\Delta T$
and mean molecular weight per electron to $\mu_e+\Delta\mu_e$ at the top of the convection
zone, where $T_i$ and $\mu_e$ are the values of these quantities within the isothermal
zone. We assume that $\mu_i$ does not change appreciably.
Note that enhanced neutronization in the convection implies $\Delta\mu_e>0$.
We set both $P$ and $\rho$ to be continuous at the convective boundary,
and expand to first order,
\be
	\frac{\Delta T}{T_i} = \frac{4}{3}\frac{\mu_i m_p}{k_{\rm B}T_i}K\frac{\rho^{1/3}}{\mu_e^{4/3}}
		\frac{\Delta\mu_e}{\mu_e}.
\ee
Recognizing that $E_{\rm F}= 4K(\rho/\mu_e)^{1/3}m_p$, this can be written more conveniently
as
\be
	\frac{\Delta T}{T_i} = \frac{1}{3}\frac{\mu_i}{\mu_e}\frac{E_{\rm F}}{k_{\rm B}T_i}
		\frac{\Delta\mu_e}{\mu_e}
	= \frac{1}{3}\frac{\mu_i}{\mu_e}\frac{E_{\rm F}}{k_{\rm B}T_i}\frac{\Delta Y_e}{Y_e},
\ee
where we have defined $\Delta Y_e$ to be positive
(i.e., $\mu_e+\Delta \mu_e=Y_e-\Delta Y_e$).
Setting $E_{\rm F}=1.9\ {\rm MeV}(2/\mu_e)^{1/3}\rho_8^{1/3}$, where
$\rho_8=\rho/10^8\ {\rm g\ cm^{-3}}$, this is rewritten as
\be
	\frac{\Delta T}{T_i} = 5.0\times10^2
	\frac{\rho_8^{1/3}}{T_{i,8}}
	\lp\frac{\mu_i}{13.7}\rp\lp\frac{2}{\mu_e}\rp^{4/3}
	\frac{\Delta Y_e}{Y_e},
\ee
where $T_{i,8}=T_i/10^8\ {\rm K}$ and $\mu_i=13.7$ for a plasma with $X_{12}=X_{16}=0.5$.
The large prefactor demonstrates that
degeneracy greatly enhances the small changes in $\mu_e$ on the temperature
discontinuity. This is because the ions provide only a small contribution to the pressure,
so that a large temperature jump is needed to offset a small change in
density. Using equation (\ref{eq:deltamc}), we find
\be
	\Delta M_c = -0.09\ M_\odot 
	\frac{\rho_8^{1/3}}{T_{c,8}/8}\lp\frac{\mu_i}{13.7}\rp\lp\frac{2}{\mu_e}\rp^{10/3}
	\frac{\Delta Y_e/Y_e}{10^{-3}},
	\label{eq:mcmue}
\ee
where $T_{c,8}=T_c/10^8\ {\rm K}$. It is interesting to note that
although $T_i$ affects the absolute
location of $M_c$ (see eq. [\ref{eq:mct}]), $\Delta M_c$ is in fact {\it independent}
of $T_i$.

\subsection{Mean Molecular Weight per Ion  and\\ Coulomb Correction Discontinuities}
\label{sec:nuclei}

   We next consider the temperature difference from changes in the composition
of the nuclei. This is of relevance because
evolution of the progenitor star during the helium burning stage enhances
$^{12}$C versus $^{16}$O with larger radii \citep{str03}.

   First we present the effect of an increase of the mean molecular weight in the convection
zone, $\Delta\mu_i$, using the equation of state given by equation (\ref{eq:eos}).
Setting the density and pressure continuous across the convective boundary,
\be
	\frac{\Delta T}{T_i}=\frac{\Delta\mu_i}{\mu_i}.
\ee
As a concrete example, we consider the composition in Model 3 from Table
\ref{tab:models}, which gives $\Delta\mu_i/\mu_i\approx0.06$. This
implies a change in convective mass
of
\be
	\Delta M_c = -1.1\times10^{-2}\ M_\odot
	\frac{\rho_8^{1/3}}{T_{c,8}/8}
	\lp\frac{2}{\mu_e}\rp^2
	\frac{\Delta\mu_i/\mu_i}{0.06},
	\label{eq:mcmui}
\ee
which is negligible.

\begin{figure}
\epsscale{1.15} 
\plotone{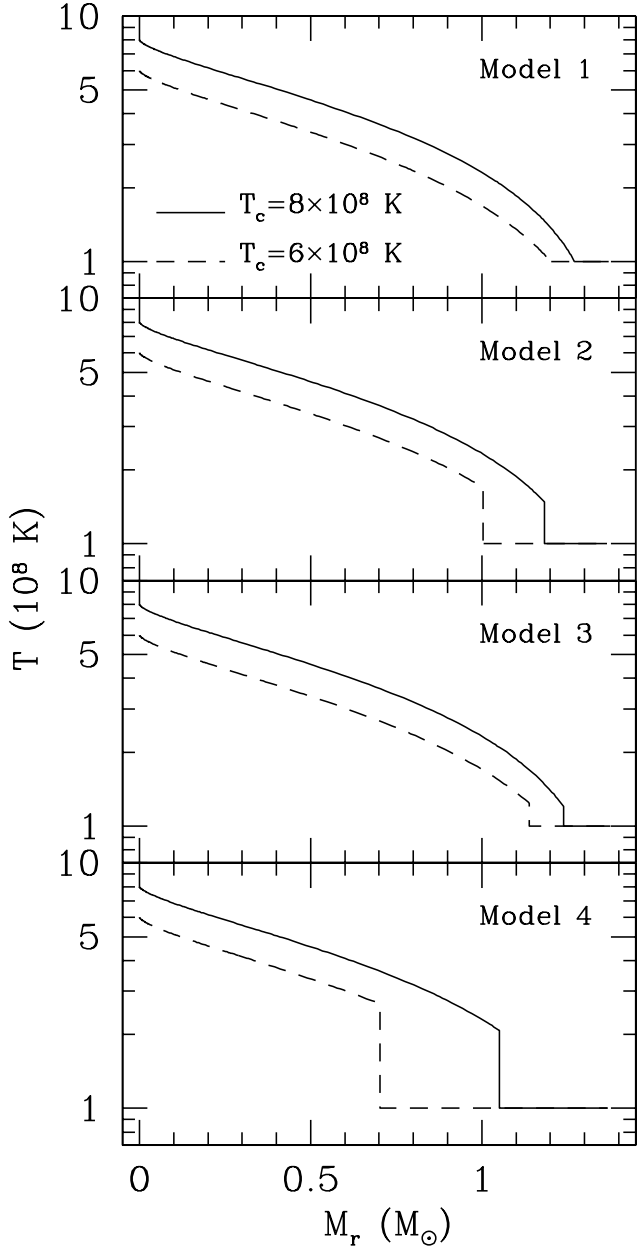}
\caption{Temperature profiles for the compositional models summarized in Table
\ref{tab:models}. For each model we consider a central temperature of
$T_c=6\times10^8\ {\rm K}$ ({\it dashed lines}) and $T_c=8\times10^8\ {\rm K}$
({\it solid lines}). All models have an isothermal temperature of $T_i=10^8\ {\rm K}$
and a mass of $1.37\ M_\odot$.
The convective boundary is at the temperature break.}
\label{fig:boundary}
\epsscale{1.0}
\end{figure}

   However, compositional changes also alter the Coulomb contributions to the internal
energy. For a multi-component plasma, the strength of this effect is measured via
the Coulomb parameter
\be
	\Gamma=\frac{\langle Z^2\rangle e^2}{ak_{\rm B}T}
		=  21.1\frac{\rho_8^{1/3}}{T_8}\frac{\langle Z^2\rangle}{48}\lp\frac{13.7}{\mu_i}\rp^{1/3},
\ee
\citep{sch99} where $a$ is the mean ion separation defined as
$4\pi a^3\rho/(3\mu_i m_p)=1$ and
\be
	\langle Z^2\rangle =  \mu_i\sum_i \frac{Z_i^2}{A_i}X_i.
\ee
We take the fitting function found by \citet{cp98} in the
limit $\Gamma\gg1$ to model Coulomb effects,
\be
	P = K\lp\frac{\rho}{\mu_e}\rp^{4/3}
		+\frac{\rho k_{\rm B}T}{\mu_i m_p}\left[1-\frac{A_1}{3}\Gamma \right],
\ee
where $A_1 = 0.9052$ is a fitting parameter. To isolate the compositional and temperature
dependencies of $\Gamma$, we write $\Gamma=\Gamma_0/T$.
Within the convection, $\Gamma$ is larger by a fractional amount
\be
	\frac{\Delta\Gamma}{\Gamma}=\frac{\Delta\Gamma_0}{\Gamma_0}-\frac{\Delta T}{T_i},
\ee
where $\Delta\Gamma_0$ is strictly from changes in composition (i.e., changes in
$\langle Z^2\rangle$ and $\mu_i$). Setting the pressure to be continuous across the convective
boundary we find
\be
	\frac{\Delta T}{T_i}= \frac{A_1}{3}\Gamma\frac{\Delta\Gamma_0}{\Gamma_0}.
\ee
Due to the factor of $\Gamma$, a small fractional change $\Delta\Gamma_0/\Gamma_0$
of merely $\sim10\%$ (as present for Model 3) implies a mass change of
\be
	\Delta M_c = -0.12\ M_\odot
	\frac{\rho_8^{1/3}}{T_{c,8}/8}
	\frac{\langle Z^2\rangle}{48}
	\lp\frac{13.7}{\mu_i}\rp^{1/3}\lp\frac{2}{\mu_e}\rp^{2}
	\frac{\Delta\Gamma_0/\Gamma_0}{0.1}.
	\nonumber
	\\
	\label{eq:mccoulomb}
\ee
Comparing equations (\ref{eq:mcmue}), (\ref{eq:mcmui}), and (\ref{eq:mccoulomb}) shows
that both changes in $Y_e$ and in Coulomb corrections can make non-negligible corrections
to the convective boundary. Which effect is largest depends on the specific
progenitor model.

\begin{figure}
\epsscale{1.15} 
\plotone{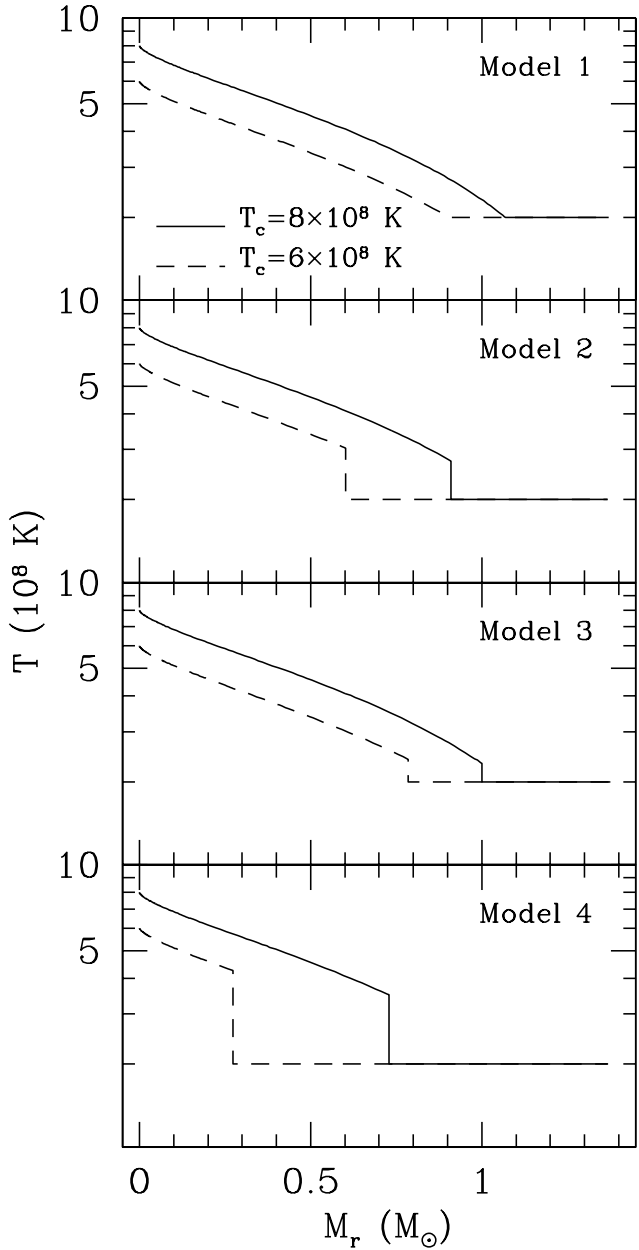}
\caption{The same as Fig. \ref{fig:boundary}, but with $T_i=2\times10^8\ {\rm K}$.}
\label{fig:boundary2}
\epsscale{1.0}
\end{figure}

\subsection{Numerical Models}
\label{sec:numerical}

   We now calculate a few models to compare how the convective
boundary changes due to compositional discontinuities. These models use
the same microphysics as described in \S \ref{sec:luminosity}. Each model
is summarized in Table \ref{tab:models} and is motivated by a plausible
progenitor scenario. Model 1 does not have a compositional discontinuity
and is included for purposes of comparison. Model 2 shows the effects
of neutronization by having a $\Delta Y_e\approx8\times10^{-4}$,
comparable to the level of neutronization found by \citet{pb07}.
For simplicity we assume that all the additional
neutrons are in $^{13}$C (instead of heavier elements such as $^{23}$Ne).
Model 3 considers a change in the mass fractions of $^{12}$C and $^{16}$O
to show the effects of a compositional gradient. Finally, Model 4 shows the
maximum effect of sedimentation by assuming that all of the $^{22}$Ne has
sunk into the convective core.

   In Figure \ref{fig:boundary} and \ref{fig:boundary2}
we compare the four models for central temperatures
of $T_c=6\times10^8\ {\rm K}$ and $8\times10^8\ {\rm K}$, and
a WD mass of $1.37\ M_\odot$. In Figure \ref{fig:boundary} we set
$T_i=10^8\ {\rm K}$, while in Figure \ref{fig:boundary2} we set $T_i=2\times10^8\ {\rm K}$.
While our models are not self-consistent in that we assume a
fixed, non-evolving compositional discontinuity, they show that the location of the convective boundary
can vary substantially depending on the progenitor
model and the nuclear reactions that take place during the simmering.
Comparing Figures \ref{fig:boundary} and \ref{fig:boundary2} shows that
$\Delta M_c$ can be large even when $T_i$ is increased. This is roughly consistent
with the analytics of \S\S \ref{sec:neutronization} and \ref{sec:nuclei} that shows
$\Delta M_c$ is independent of $T_i$.


\section{Conclusion and Discussion}
\label{sec:conclusion}

   We have highlighted some important properties of the convection
present during the pre-explosive carbon simmering phase of SNe Ia. The convective
velocities near the top boundary are decreased significantly because the convective
luminosity is extracted to heat and grow the convective zone and the long conductive
timescale of the non-convective exterior enforces $L_c = 0$ at the top of the
convective zone. The size of the convective zone can change depending on
compositional gradients within the progenitor WD.

   Perhaps the most severe effect depends on whether there is time available
for sedimentation of $^{22}$Ne prior to carbon ignition. There remains considerable
uncertainty in the timescale for this process to occur. For a $1.2M_\odot$ WD, complete
sedimentation requires $\sim9\ {\rm Gyr}$ \citep{db02} with more massive WDs having
an even shorter sedimentation time, but this all depends sensitively on the size
of the diffusion coefficient used. The significant impact this may have on SNe Ia
progenitors means that better theoretical calculations of this diffusion coefficient is an
essential concern.
The work by \citet{dal06} is an important step in this direction, but the multi-component
plasma present in the WD may be a crucial detail.

The presence of the convective core prior to explosive carbon burning may have
other implications for SNe Ia that deserve a closer look.  The convective motions
should give rise to a massive core dynamo, which setting $B^2/8\pi\sim\rho V_c^2$
implies fields on the order of $B\sim10^{12}\ {\rm G}$. In addition, the motion of these
convective eddies may stochastically excite waves
(both {\it g}-modes and {\it p}-modes). These waves propagate away from the
convective zone and transfer energy to shallower regions. Both of these processes
may prove important in understanding the evolution toward explosive burning, and we
plan to investigate them in more detail in a future study.

 Multidimensional simulations are needed to study the physics
of this simmering phase not captured by our simple calculation. These simulations are very challenging in that the flow
must be followed for many turnover times. As a result, implicit Eulerian schemes \citep[see for instance
in two dimensions of][]{hs02,sw06},
anelastic codes \citep{kwg06} or a low Mach number formulation
\citep{alm06a,alm06b} are required. These simulations will be helpful in characterizing the flow
topology \citep[e.g.,][]{kwg06} and the number and spatial distribution
of ignition points.

The simmering phase helps set the nature of the eventual runaway and
subsequent explosion. \citet{woo04} and \citet{ww04}
demonstrate that the summering is paramount for setting the initial
ignition points for explosive burning.  The large scale structure of
the convective turbulence may also be important for flame propagation. For instance,
\citet{hs02} argue that the initial velocities
of the burning fronts are determined by the background convective motions
and not by the laminar flame speed or Rayleigh-Taylor instabilities. This suggests that the initial explosive burning is likely to be off-centered
\citep{gw95,woo04}.
The background turbulent state will also likely subsequently affect the
motion of the bubbles by acting as a viscous drag \citep{zd07}.

Finally, the material above the convective core is devoid of this turbulence. The
burning properties may change in an interesting manner as the flame passes into
this relatively "quiet" region.
Some have argued that a delayed detonation transition (DDT) of the burning
may be needed to match observations \citep{ple04,liv05}. Although the
concept of a DDT has been considered for some time \citep{kho91,ww94},
how and if it occurs is still uncertain \citep{nw97,nie99,woosley07}. The position
of the convective boundary may be an important detail. At the end of simmering it
typically lies at a density of $\sim3\times10^7-10^8\ {\rm g\ cm^{-3}}$, which may become
near the density usually invoked for a DDT ($\sim10^7\ {\rm g\ cm^{-3}}$)
once the WD has expanded from the propagation of the deflagration wave.

\acknowledgements
We thank Lars Bildsten, Eliot Quataert, Nevin Weinberg, and Stan Woosley for
helpful discussions. P. C. is supported by the Miller Institute for
Basic Research.


\end{document}